%% file: mspes.tex
\documentclass[preprint]{aastex}
\usepackage[colorlinks=true,urlcolor=black,linkcolor=black,citecolor=black]{hyperref} 
\usepackage{natbib}
\usepackage{epsfig}

\shorttitle{Polarimetric Emission survey}
\shortauthors{Mitra et al.}
\begin{document}
\title{Meterwavelength Single-pulse Polarimetric Emission Survey} 

\author{Dipanjan Mitra\altaffilmark{1,2,3}, Rahul Basu\altaffilmark{3}, Krzysztof Maciesiak\altaffilmark{3},
Anna Skrzypczak\altaffilmark{3}, George I. Melikidze\altaffilmark{3,4}, Andrzej Szary\altaffilmark{5,3}, 
Krzysztof Krzeszowski\altaffilmark{3}} 

\altaffiltext{1}{National Centre for Radio Astrophysics, Ganeshkhind, Pune 411 007, India}
\altaffiltext{2}{Physics Department, University of Vermont, Burlington VT 05405}
\altaffiltext{3}{Janusz Gil Institute of Astronomy, University of Zielona G\'ora, ul. Szafrana 2, 65-516 Zielona G\'ora, Poland}
\altaffiltext{4}{Abastumani Astrophysical Observatory, Ilia State University, 3-5 Cholokashvili Ave., Tbilisi, 0160, Georgia}
\altaffiltext{5}{ASTRON, the Netherlands Institute for Radio Astronomy, Postbus 2, 7990 AA, Dwingeloo, The Netherlands}
\email{dmitra@ncra.tifr.res.in}

\begin{abstract}
\noindent
We have conducted the Meterwavelength Single-pulse Polarimetric Emission 
Survey to study the radio emission properties of normal pulsars.
A total of 123 pulsars with periods between 0.1 seconds and 8.5 
seconds were observed in the survey at two different frequencies, 
105 profiles at 333 MHz, 118 profiles at 618 MHz and 100 pulsars at both. 
In this work we concentrate primarily on the time-averaged properties 
of the pulsar emission. The measured widths of the pulsar profiles 
in our sample usually exhibit the radius to frequency mapping.
We validate the existence of lower bounds for the distribution of profile 
widths with pulsar period ($P$), which is seen for multiple definitions 
of the width, viz. a lower boundary line (LBL) at $2.7^{\circ} P^{-0.5}$ with 
width measured at 50\% level of profile peak, a LBL at $5.7^{\circ} P^{-0.5}$ 
for 10\% level of peak and LBL at $6.3^{\circ} P^{-0.5}$ for width defined 
as 5$\sigma$ above the baseline level. In addition we have measured the 
degree of linear polarization in the average profile of pulsars and confirmed 
their dependence on pulsar spindown energy loss ($\dot{E}$). The single pulse 
polarization data show interesting trends with the polarization position angle 
(PPA) distribution exhibiting the simple rotating vector model for high 
$\dot{E}$ pulsars while the PPA becomes more complex for medium and low 
$\dot{E}$ pulsars. The single pulse total intensity data is useful for 
studying a number of emission properties from pulsars like subpulse drifting, 
nulling and mode changing which is being explored in separate works.
\end{abstract}


\section{INTRODUCTION}
The coherent radio emission from pulsars are broadband in nature
ranging typically from tens of MHz to a few GHz.  The pulsar
population can be categorised into two distinct groups based on their
rotation periods $P$ and period derivatives $\dot{P}$, namely the
millisecond pulsars (MSP) with $P \lesssim 30$ milliseconds and $\dot{P} <
10^{-19}$ s/s and normal pulsars with $P \gtrsim 100$ milliseconds and
$\dot{P} \gtrsim 10^{-17}$ s/s. In our present work we focus primarily
on the normal pulsars.  The coherent radio emission from pulsars are
believed to originate as a result of the growth of instabilities in
the relativistic plasma streaming along super-strong magnetic field
lines within the pulsar magnetosphere
(e.g. \citealt{1995JApA...16..137M}).  The physical processes
that lead to the radio emission in pulsars are topics of active
research and their understanding would greatly benefit from new
observational insights.

The individual pulses of a radio pulsar are highly variable and have
proved essential in understanding various aspects of pulsar emission
mechanism.  
The variations in the total intensity single pulses reveal interesting 
features like microstructures, giant pulse, mode changing, nulling and 
subpulse drifting in the radio emission. These effects provide insights 
into the dynamics of the non-stationary processes in the pulsar magnetosphere 
at short timescales, ranging from sub-nanoseconds to a few seconds. Extreme 
examples are the nanosecond pulses reported from the Crab pulsar
(e.g \citealt{2007ApJ...670..693H}, \citealt{2010A&A...524A..60J}) and 
in PSR B1937+21 (\citealt{2004ApJ...616..439S}). 
The single pulse polarization
data showed the presence of orthogonal polarization modes (OPM) and
circular polarization which provide important clues about the
emergence and propagation of the emission modes within the
magnetospheric plasma (e.g. \citealt{2009ApJ...696L.141M},
\citealt{2014ApJ...794..105M}).  The time averaged properties of pulsar
emission (known as the pulsar profile), obtained after averaging
individual pulses over several tens of minutes, are observed to be
highly stable which is indicative of the global properties of the
pulsar magnetosphere.  The change of the profile width as a function
of frequency reflect the radius to frequency mapping (RFM) likely
signifying the radio emission at different frequencies originating at
different heights (e.g. \citealt{2002ApJ...577..322M}).  
The polarization position angle (PPA) of the
linear polarization in the average profile resembles a characteristic
S-shaped traverse which according to the rotating vector model (RVM,
\citealt{1969ApL.....3..225R}) is interpreted as radiation arising
from regions of dipolar magnetic field lines and often used to
estimate the pulsar geometry.  The pulsar profile is also important
for determining the shape of emission beam and location of the radio
emission within the magnetosphere (e.g. \citealt{1993ApJ...405..285R},
\citealt{1999A&A...346..906M}). It is evident that the complete
characterization of the pulsar emission is accomplished by a
multifrequency ($\sim$ 100 MHz to 5 GHz) approach involving a thorough
understanding of the average profile as well as single pulse nature in
both total intensity and polarimetric data in the wider pulsar
population, exemplified by a host of studies
\citet{1975ApJ...195..513T}, \citet{1975ApJ...196...83M},
\citet{1988MNRAS.234..477L}, \citet{1991ApJ...370..643B},
\citet{1997A&AS..126..121V}, \citealt{1998MNRAS.301..235G} (GL98
hereafter), \citealt{2008MNRAS.388..261J} (JKMG08 hereafter),
\citet{1984ApJS...55..247S}, \citet{2011ApJ...727...92M}.
Pulsars have also been detected at millimeter wavelengths, 
with a tentative evidence for the possible turn up in the spectra
which gives additional clues for the emission changing from
coherent to incoherent mode (e.g. \citealt{1996A&A...306..867K}).

A number of studies involving the time averaged total intensity and
polarization properties of pulsars have been reported in the
literature, notably the comprehensive studies by GL98 at five
frequencies between 234 MHz and 1640 MHz using the Lovell telescope
and JKMG08 at multiple frequencies between 234 MHz and 3100 MHz using
the Parkes Telescope and the Giant Meterwave Radio Telescope (GMRT).
Polarization observations below 200 MHz are rare with the most notable
exception of the recent study by \citet{2015A&A...576A..62N} using the Low Frequency
Array (LOFAR).  Higher frequency polarization studies have used the
Effelsberg radio telescope (\citealt{1998ApJ...501..286X},
\citealt{1998A&A...334..571V}) and the Parkes Telescope
(\citealt{2005MNRAS.359..481K}, \citealt{2006MNRAS.368.1856J}).  The
results have been utilized in constructing the shape of pulsar
emission beam (e.g. \citealt{1999A&A...346..906M},
\citealt{2007MNRAS.380.1678K}), estimating the radio emission heights
(e.g. \citealt{2004A&A...421..215M}, \citealt{1997MNRAS.288..631K})
and probing the validity of the RFM \citep{2002ApJ...577..322M}.

Previous single pulse surveys have mostly concentrated on the
total intensity observations with a number of significant contributions aimed at
understanding the phenomenon of mode changing, subpulse drifting and
nulling (e.g. \citealt{2012MNRAS.423.1351B},
\citealt{2006A&A...445..243W}, \citealt{2007A&A...469..607W},
\citealt{2007MNRAS.377.1383W}).  In comparison single pulse
polarization studies are relatively sparse.  Some of the systematic
single pulse polarization studies have been conducted using the
Arecibo and the GMRT at 325 MHz and 1400 MHz
\citep{1984ApJS...55..247S,2011ApJ...727...92M,2015ApJ...806..236M}.
A few sporadic studies using other telescopes involve mostly the
brightest pulsars, e.g. Lovell: \citet{1995MNRAS.276L..55G};
Westerbork: \citet{2004A&A...421..681E}, Parkes:
\citet{2001ApJ...549L.101J}, Effelsberg and Westerbock:
\citet{2002A&A...391..247K}.  These observations revealed important
insights like the existence of two OPMs which follow the RVM
\citep{1995MNRAS.276L..55G}, the nature of partial cone emission
\citep{2011ApJ...727...92M}, existence of polarization microstructure
\citep{2015ApJ...806..236M}, etc.

In the light of the above discussion it is apparent that there is a
shortage of high quality single pulse polarization data for the wider
pulsar population.  The primary objective of this work is to conduct a
survey of single pulse polarization in a large sample of pulsars for
studying multiple aspects of pulsar radio emission.  We report two
frequency 333 and 618 MHz GMRT single pulse polarimetric observations
of 123 pulsars lying in the declination range -50$^\circ$ to
+25$^\circ$.  For a majority of these pulsar single pulse polarization
data could be obtained.  In section~\ref{sec2} and \ref{sec3} we
discuss the sample selection, the observational detail and data
analysis procedure.  We present the results and discussion of the
time-averaged and single pulse properties in section~\ref{sec4} and
summarize the results in section~\ref{sec5}.

\section{SAMPLE SELECTION}
\label{sec2}
 
In recent years a large number of pulsars have been discovered by the
Parkes radio telescope (\citealt{2004MNRAS.352.1439H}) in the declination range south of +25$^\circ$
which are largely inaccessible to the majority of radio telescopes
located at higher northern latitudes. This has resulted in a dearth of
single pulse polarization data for most of these pulsars particularly
at meter wavelengths. The GMRT operating at meter wavelengths and
located at relatively low latitudes is suited for these studies in
pulsars located north of -50$^\circ$ declination.  The GMRT is one of
the most sensitive radio telescope at meter wavelengths, second only
to the Arecibo radio telescope in terms of sensitivity (the collecting
area of full GMRT is 0.67 times that of the Arecibo telescope), but
with a greater sky coverage (Arecibo telescope can observe in the
declination range between 0$^\circ$ and +35$^\circ$).

We have selected our sample from the ATNF pulsar database\footnote{\url{http://atnf.csiro.au/research/pulsar/psrcat/}} 
\citep{2005AJ....129.1993M} to be observed with the GMRT at 
333 and 618 MHz in the declination range  -50$^\circ$ to +25$^\circ$
\footnote{ The current single pulse study should be contrasted with
the time-averaged polarization work by JKMG08 using GMRT with
similar declination coverage.  However, their experiment was
designed to observe 3 frequencies simultaneously, using different
sub-arrays.  This resulted in lower sensitivities due to lesser number
of antennas used at each frequency.  In addition their sensitivities
were also diminished due to a smaller frequency bandwidth
coverage.}.  The selection criterion were as follows; we restricted
the sources to dispersion measures (DM) lower than 200~pc~cm$^{-3}$
primarily to avoid scattering.  However, some of the pulsars with DM
$>$ 150~pc~cm$^{-3}$ were affected by scattering at 333 MHz, but
even in these cases the 618 MHz data remained unaffected and suitable
for our studies.  In addition we only selected pulsars with estimated
flux larger than 5 mJy at 618 MHz.  This was motivated by our desire
to study single pulses with a signal to noise ratio (S/N) in excess of
10 using the GMRT.  The selection criterion yielded 123 pulsars in the
period range of $0.1$ seconds to $8.5$ second.  The majority of the
pulsars have no previous single pulse polarization data, but we have
also included a few well studied pulsars in our sample for calibration
and verifying our analysis schemes.  Table~\ref{tab3} lists the
pulsars in our sample.

\section{OBSERVATIONS}
\label{sec3}

The pulsar observations were carried out at the GMRT between January
and August 2014 covering 25 observing days and roughly 180 hours Telescope time.  
The GMRT is an interferometer consisting of 30 antennas, each of 45 meter diameter, 
operating at six different frequencies between 150 MHz and 1450 MHz
\citep{1991CuSc...60...95S}.  The antennas are arranged resembling a
Y-shaped array with two distinct configurations, a central square
populated by 14 antennas within an area of 1 square km, and the
remaining 16 antennas spread out along three arms over a 25 km
diameter.  Pulsar observations are usually conducted in the
phased-array (PA) mode (see \citealt{2000ASPC..202..277G};
\citealt{2000Phd}) where the signals from different antennas are
co-added in phase.  In order to reach sufficient sensitivity for
single pulse studies we used around 20 antennas in the phased-array
which included all the available antennas in the central square and
the two nearest arm antennas.  The extreme arm antennas were not
included since they would dephase very fast (within 15 minutes), as a
result of ionospheric variations, and would reduce the effective S/N.

We observed at two separate frequency bands roughly between 317$-$333
MHz and 602$-$618 MHz.  At these observing frequencies, the GMRT is
equipped with dual linear feeds which are converted to left and
right-handed circular polarizations via a hybrid.  The dual
polarization signals are passed through a superheterodyne system and
down-converted to the baseband which are finally fed to the GMRT
software backend (GSB, \citealt{2010ExA....28...25R}).  The FX
correlator algorithm is implemented in the GSB where each polarization
voltage is first digitally sampled at the nyquist rate and
subsequently fast fourier transformed to produce frequency channels
across the bands which are then cross correlated.  In the PA mode of
the GSB, with full Stokes capability, the voltage signals from all the
selected antennas are added for each polarization to get dual
polarization data which are used to produce the raw Stokes values
using a multiplying polarimeter algorithm.  We used a total bandwidth
of 16.67 MHz spread over 256 channels with time resolution of 0.245
milliseconds.  The final time-series data were equivalent to a large single
dish and similar polarization calibration can be applied to get the
calibrated Stokes parameters
(\citealt{2002PASA...19..277J,2005URSI...J03a,2007MNRAS.379..932M},
JKMG08).

Our polarization calibration and analysis were similar to JKMG08.  The
antennas were initially phase aligned with respect to a reference
antenna using a strong flux calibrator as a model.  Before recording
each pulsar a secondary phase calibrator close to the source was used
to verify the phase alignment of the antennas and the phases realigned
if needed.  The auto and cross-polarized data were recorded in a
filterbank format and were analysed using the software schemes
developed by \citet{2005URSI...J03a} and later used in JKMG08.  The
polarization data were gain corrected and converted to the four Stokes
parameters ($I,Q,U,V$) for each individual spectral channels.  Proper
corrections were applied for the fixed delay between the circular
polarization channels of the reference antenna.  This was followed by
correcting for the phase lags caused by Faraday rotation in the
interstellar medium using the known rotation measure (hereafter RM) of the pulsar
(ATNF database, also given in table~\ref{tab2}). We further verified the catalog rotation
measures by using the technique
of maximizing the linear polarization across the observing 
band (e.g. \citealt{2003A&A...398..993M},
\citealt{2005A&A...441.1217B}, \citealt{2015MNRAS.451.2493S}).
Using our analysis methods we were not able to measure RM with any accuracy for these cases
and used RM value of 0.0 $rad/m^2$ for analysis. In one pulsar J1321+8323 we were 
able to estimate from our data the previously unreported RM. 
Finally, we searched for and rejected spectral
channels affected by radio frequency interference (RFI) before
averaging the channels, adjusted to the upper edge of the band at 333
MHz and 618 MHz, respectively, for each of the four Stokes parameters.
During each observing run the data quality was frequently monitored by
interspersing the test pulsars B0950+08, B1133+16 and B1929+10 at
regular intervals and different parallactic angles.  We have estimated
the systematic error in the linear and circular polarization measurements 
to be $\sim$5 percent. 
\begin{figure*}
\begin{center}
\begin{tabular}{c}
{\mbox{\includegraphics[angle=0,scale=0.5]{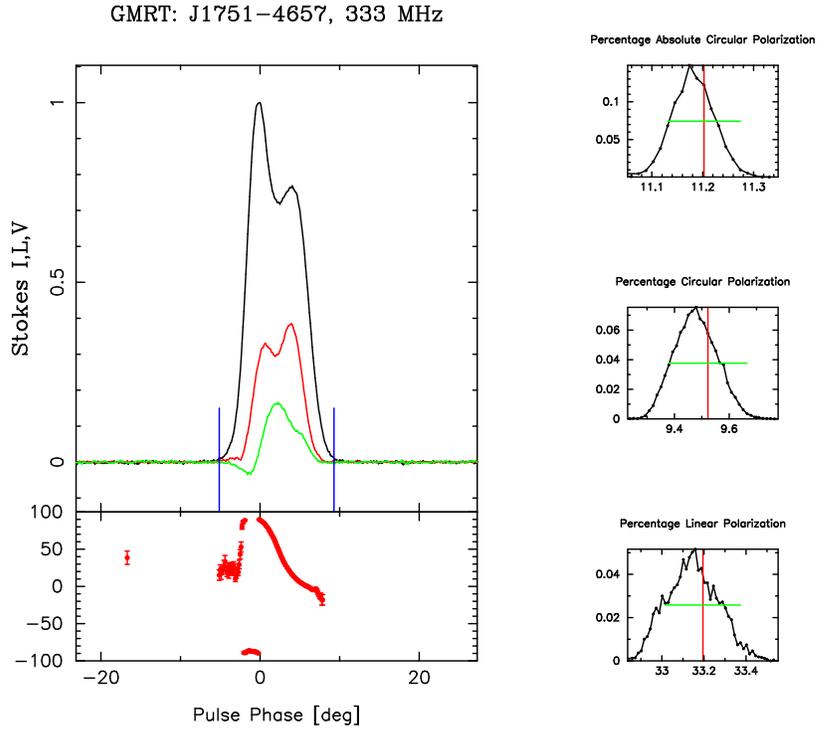}}}\\
\end{tabular}
\caption{The left plot shows the time-averaged polarisation properties of PSR J1751-4657 at 333 MHz. The three 
panels in the right shows the distribution of the variation of $\%L$, $\%V$ and $\mid V \mid$ (from bottom to top) 
that arise due to the noise in the baseline. The red line shows the median of the distribution and the 
green line shows the rms. See section~\ref{sec4_1_2} for details. The figures for all pulsars is 
available for download (see appendix).} 
\label{fig1}
\end{center}
\end{figure*}

\begin{figure*}
\begin{center}
\begin{tabular}{r}
{\mbox{\includegraphics[angle=0,scale=0.72]{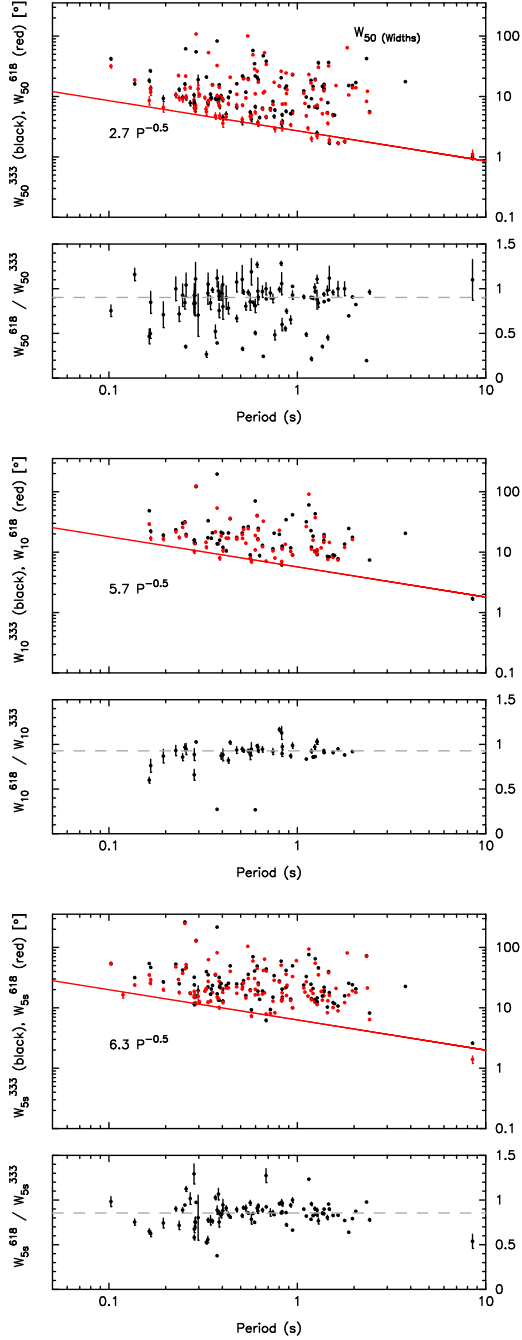}}}\\
\end{tabular}
\caption{The three panels shows W$_{50}$, W$_{10}$ and $W_{5\sigma}$ width in our sample 
as given in table~\ref{tab3} as a function of pulsar period. 
The black and red points correspond to 333 MHz and 618 MHz respectively. 
The red lines in the figures correspond to 2.7$^{\circ}P^{-0.5}$, 5.7$^{\circ}P^{-0.5}$ and 6.3$^{\circ}P^{-0.5}$ respectively. 
The bottom panel shows the ratio of the widths at the two frequencies as a function of pulsar period. 
The grey dashed lines are at 0.90, 0.92 and 0.84 respectively and shows the median values of the ratio. 
See section~\ref{sec4_1_2} for details.} 
\label{fig2}
\end{center}
\end{figure*}

\begin{table}[h]
\caption[]{MSPES Observational Summary:} 
    \begin{tabular}{cccccccc}
        \hline
Freq & Total BW & Channel BW & t$_{\rm res}$ & N$_{\rm TAP}$ & N$_{\rm SP}$ & N$_{\rm pol}$ & N$_{\rm Drift}$ \\
(MHz)&  (MHz)   &  (MHz)     &  (msec)       &               &              &               &                 \\
        \hline
      &         &        &        &      &    &    &    \\
 333  & 16.66   & 0.065  & 0.245  & 105  & 83 & 59 & 39 \\
 618  & 16.66   & 0.065  & 0.245  & 118  & 93 & 49 & 44 \\
Common& $-$     &  $-$   &  $-$   & 100  & 72 & 40 & 26 \\
       \hline
    \end{tabular}
\label{tab1}\\\\
{\small The first four columns of the table specifies the observing frequency,
bandwidth, channel and time resolutions.  Column 5 gives the number of
pulsars N$_{\rm TAP}$ with good time-averaged polarization profiles,
column 6 gives the number of pulsars N$_{\rm SP}$ for which the S/N
ratio of the single pulse exceeds 5$\sigma$ limit and column 7 gives
the number of pulsars where PPA's can be measured for more than 5\% of
the single pulses.  Column 8 specifies the number of pulsars N$_{\rm
  Drift}$ that showed drifting properties (see text for details).}
\end{table}

\section{RESULTS \& DISCUSSION}
\label{sec4}
We have recorded a total of 223 good average polarization profiles
from 123 pulsars at two frequencies.  In table~\ref{tab1} we present a
short summary of the observations and in table~\ref{tab2} we 
summarize the observational results.  Our sample has 77 pulsars common
with GL98, and 30 pulsars common with JKMG08, which are the notable
polarization surveys of pulsars at similar wavelengths, though our study
had a significantly improved S/N.  Additionally 79 pulsars in our sample
were found to be common with the single pulse study of subpulse
drifting by \citet{2007A&A...469..607W} and
\citet{2006A&A...445..243W} using the Westerbork Synthesis Radio
Telescope (WSRT) at 90~cm and 21~cm.  There are 47 pulsars in our
sample that were also in the list of the pulsars used for single pulse
study by \citet{2012MNRAS.423.1351B} at 1350 MHz.

Most of the pulsars were observed for about 2100 pulses, however, in
some cases we had fewer periods due to a variety of reasons like 
presence of RFI, setting of sources during the observing run, time on
calibrator source, etc. The single pulse polarization study at 618 MHz, 
to the best of our knowledge, is the
first such survey carried out at this frequency band.  For each
observed pulsar we determined the number of usable pulses (N$_p$),
unaffected by RFI, along with the average S/N (S/N$_{\rm{avg}}$) of
the peak intensity for the entire observing run.  We further estimated
the percentage of single pulses for which the peak S/N $>$ 5 (see
table~\ref{tab2}).  We found around 77\% of the data, corresponding to
104 pulsars, useful for single pulse analysis (see table~\ref{tab1}
for full summary).  To give an assessment of the single pulse
polarization quality we quote the percentage of single pulses in which
polarization position angle (PPA) values for linear polarization S/N
$>$ 3 could be measured.  This gave around $\sim$50\% of the pulsars
which had more than 5\% of the single pulses with measured PPA values
(see table~\ref{tab1},~\ref{tab2}).  In the remainder of this
section we discuss the preliminary outcomes of our survey.
  
\subsection{Time Averaged properties}
\label{sec4_1}
The time-averaged profiles, with periods of each pulsar determined using
the $TEMPO2$\footnote{http://www.atnf.csiro.au/research/pulsar/tempo2/} software, 
were obtained by averaging the single pulses, after rejecting the ones affected by RFI (an example of
average polarization for the pulsar J1751--4657 is shown in
Fig.~\ref{fig1}).  We estimated pulse widths at each frequency ($\nu$)
using three different schemes namely the $W_{5\sigma}^{\nu}$
corresponding to the pulse width measured at five times the baseline
noise rms and $W_{10}^{\nu}$ and $W_{50}^{\nu}$ corresponding to
widths at 10\% and 50\% level of the peak intensity, respectively.
All the measured widths at each frequency are shown in
table~\ref{tab3} along with the error in pulse widths, computed using
the prescription of \citealt{1997MNRAS.288..631K} (equation 4
therein).  The average linear polarization $L(\phi_i)$ across the
profile phase, $\phi_i$, was obtained by summing up the stokes
$U(\phi_i)$ and $Q(\phi_i)$ along each $\phi_i$, and using the
relation $L(\phi_i) = \sqrt{\left(\sum_{j=1}^{n} U_j(\phi_i)\right)^2+
  \left(\sum_{j=1}^{n} Q_j(\phi_i)\right)^2}/n$, here $n$ is the total
number of pulses.  The $L(\phi_i)$ estimated above has a positive
bias, and a mean value of the linear polarization obtained from the
off pulse region is subtracted to obtain the final $L(\phi_i)$.  The
average circular polarization was obtained using the relation
$V(\phi_i)= \sum_{j=1}^{n} V_j(\phi_i)/n$.  The average polarization
position angle was obtained as $\Psi(\phi_i) =$ 0.5 tan$^{-1} \left(
\sum_{j=1}^{n} U_j(\phi_i)/\sum_{j=1}^{n} Q_j(\phi_i)\right)$, with
only points greater than three times the rms of the linear
polarization baseline level being used.

\subsubsection{\bf Pulse widths} 
\label{sec4_1_1}
Pulse width serves as a useful tool for
investigating the geometry and location of pulsar radio emission
within the magnetosphere.  In figure \ref{fig2} we show the dependence
of the different widths, W$_{50}^{\nu}$, W$_{10}^{\nu}$ and
W$_{5\sigma}^{\nu}$, on pulsar period $P$.  In table~\ref{tab3} we
have indicated 12 pulsars at 333 MHz and the pulsar J1848$-$0123 at
618 MHz which were highly scattered and not used in our plots and
statistical analysis.  The bottom panel in Fig.~\ref{fig2} shows the
ratio of the widths at the two frequencies and the dashed grey line is
the median value of the ratio. The average ratio of the three profile
measure is $\sim$0.89, and agrees with the putative phenomenon of
radius to frequency mapping (e.g. Mitra \& Rankin 2002).  Assuming a
power law dependence of the frequency on widths, $W^{\nu} \propto
\nu^a$, we find, $a \sim $-$0.19 \pm 0.1$, which agrees with previous
results
\citep{1999A&A...346..906M,2002ApJ...577..322M,2014ApJS..215...11C}.
There was one notable exception to the above results, PSR J1034--3224,
where the ratio of widths was significantly larger than unity.  On
closer inspection it was revealed that an additional emission
component appeared at 618 MHz which was absent at 333 MHz.

The pulse width decreasing with increasing $P$, seen in
Fig.~\ref{fig2}, is a well established phenomenon with a lower bound
to the distribution of widths noted by several authors
(e.g. \citealt{1988MNRAS.234..477L, 1990ApJ...352..247R,
  1993ApJ...405..285R}, GL98, \citealt{2011MNRAS.417.1444M,
  2012MNRAS.424.1762M}; \citealt{2015arXiv150906396P}).  In particular
\citet{1990ApJ...352..247R} found that $W_{50}$ for the core
components followed a lower boundary line (LBL) corresponding to $2.45^{\circ}
P^{-0.5}$ where the $P^{-0.5}$ dependence is the scaling 
of the opening angle of the open dipolar field lines 
(e.g. \citealt{1990MNRAS.245..514B},\citealt{1998ApJ...501..270K}).  
Recently \citet{2012MNRAS.424.1762M} emphasized the
existence of the same LBL, $2.45^{\circ} P^{-0.5}$, for $W_{50}$ at 1 GHz
frequency in a wider population of 1450 pulsars, which included both
core and conal components.  They argued that the LBL corresponds to
the smallest angular structures that can be observed either as core or
conal component widths.  
In Fig.~\ref{fig2} the width distribution also appears to have a lower
bound. Since the pulse widths at 618 MHz are smaller, the lower bound
is dominated by the 618 MHz measurements. We modeled the lower bound for $W_{50}$ 
by scaling the 1 GHz value of 2.45$^{\circ}$ to 618 MHz using
$\alpha\sim$-$0.19$ and find the LBL to be $2.7^{\circ} P^{-0.5}$ which
appears to be consistent with our result.  In the case of widths $W_{10}$ and 
$W_{5\sigma}$ we observed LBLs but there are no previous estimates to compare our 
results. The lower bounds, which are dominated by measurements at 618 MHz for 
$W_{10}$ and $W_{5\sigma}$, could be represented by a LBL of the form $5.7^{\circ} P^{-0.5}$ 
and $6.3^{\circ} P^{-0.5}$ respectively, and were derived by visual examination. 
Since the $W_{10}$ and $W_{5\sigma}$ are measured at comparatively 
lower intensity levels that $W_{50}$, these measurements includes both the width of 
the component as well as the separation of the components. Thus the fact that 
in these measurements the period scaling of $P^{-0.5}$ still seem to hold implies 
that the spacing between the components has a similar period scaling as the 
component ($W_{50}$) widths. The presence of a LBL in the width distribution 
is a curious phenomenon and a more detailed study to understand the LBL as a 
function of pulsar frequency and other profile measures is currently underway.

\begin{figure}
\begin{center}
\includegraphics[scale=0.62,angle=0.]{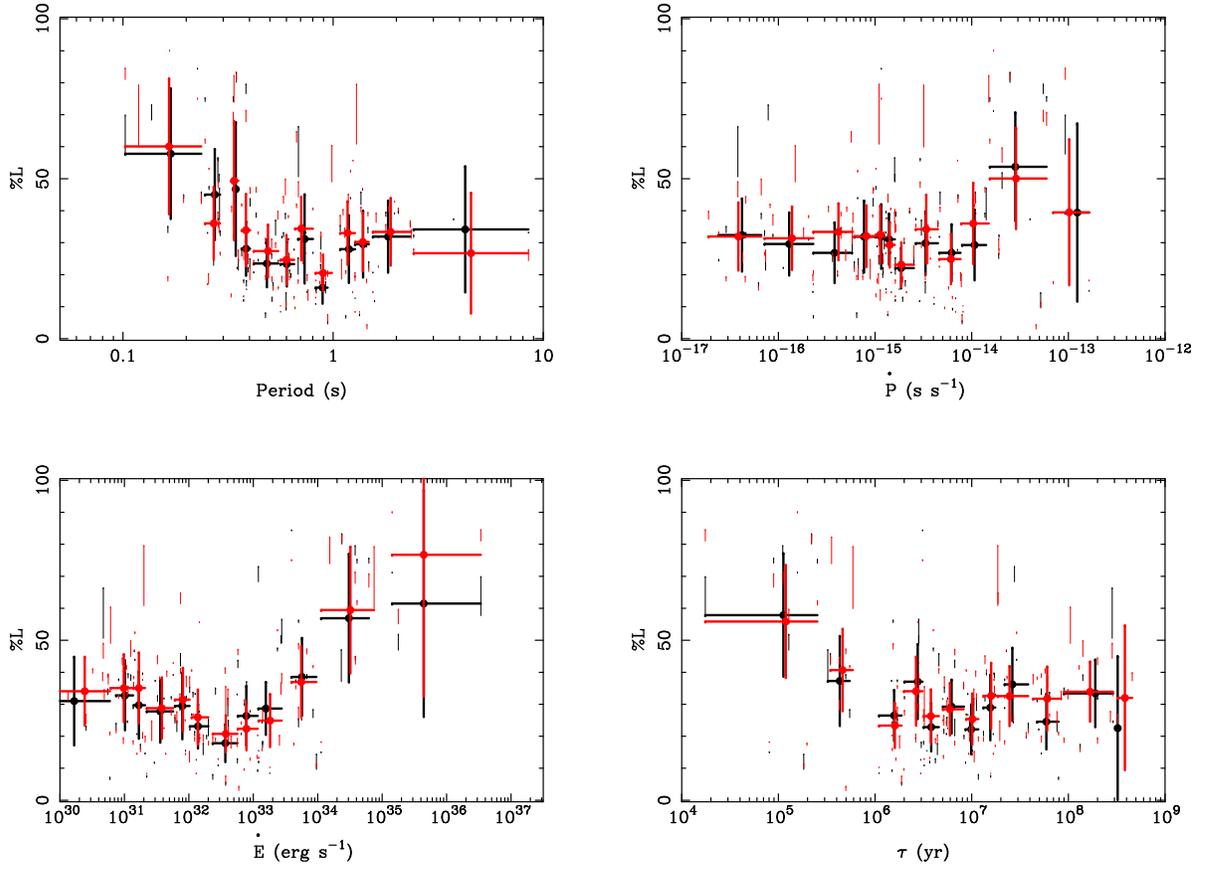}
\caption{The percentage of linear polarization for all the observed pulsars (table~\ref{tab2}) is plotted as a function of different pulsar parameters. 
The black and red short lines correspond to 333 MHz and 618 MHz respectively. 
The black and red points with error bars correspond to median values of the sample. See section~\ref{sec4_1_2} for details.
\label{fig3}}
\end{center}
\end{figure}

\begin{figure*}
\begin{center}
\begin{tabular}{cc}
{\mbox{\includegraphics[scale=0.35,angle=-0.]{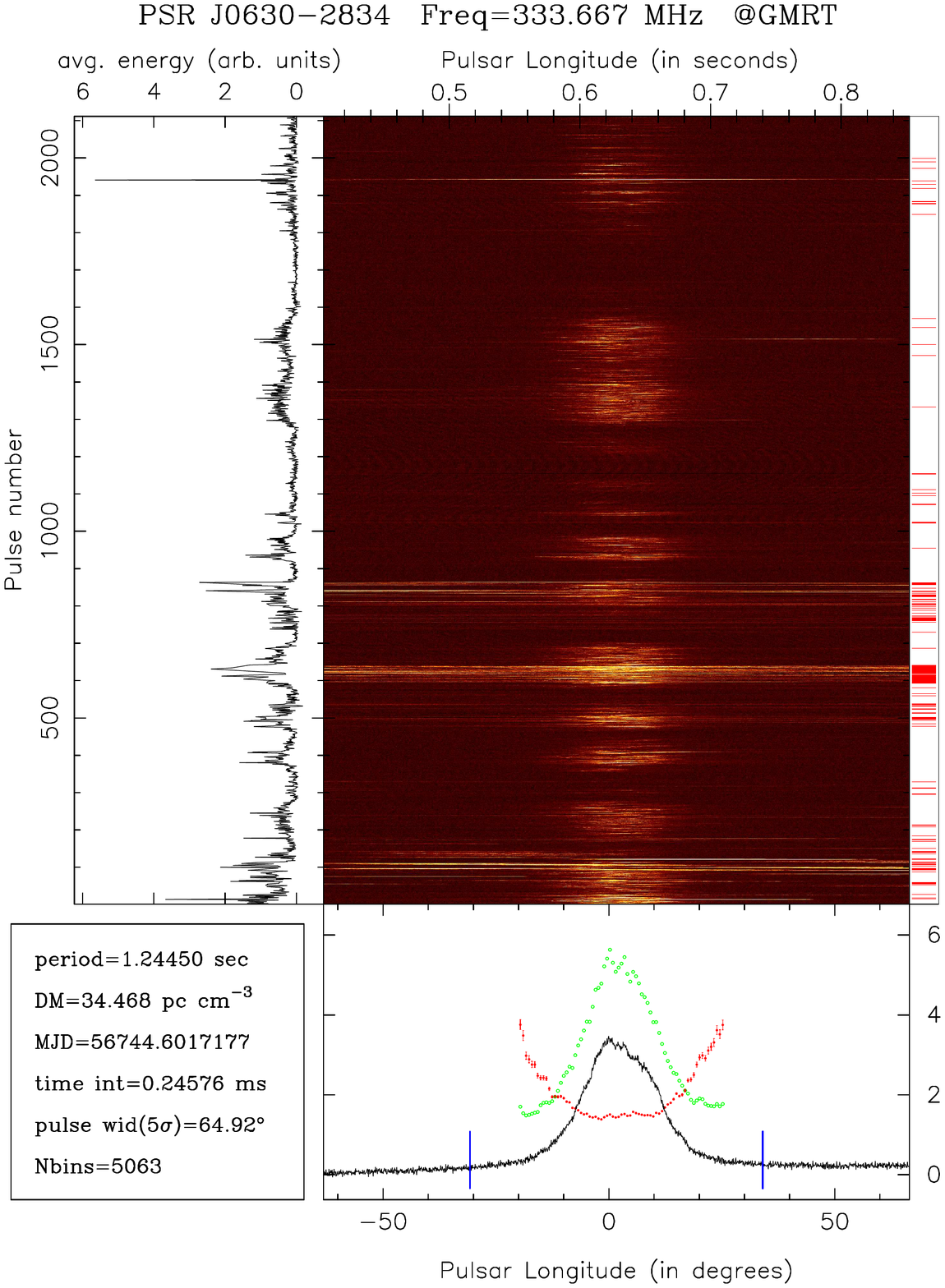}}}& 
{\mbox{\includegraphics[scale=0.35,angle=-0.]{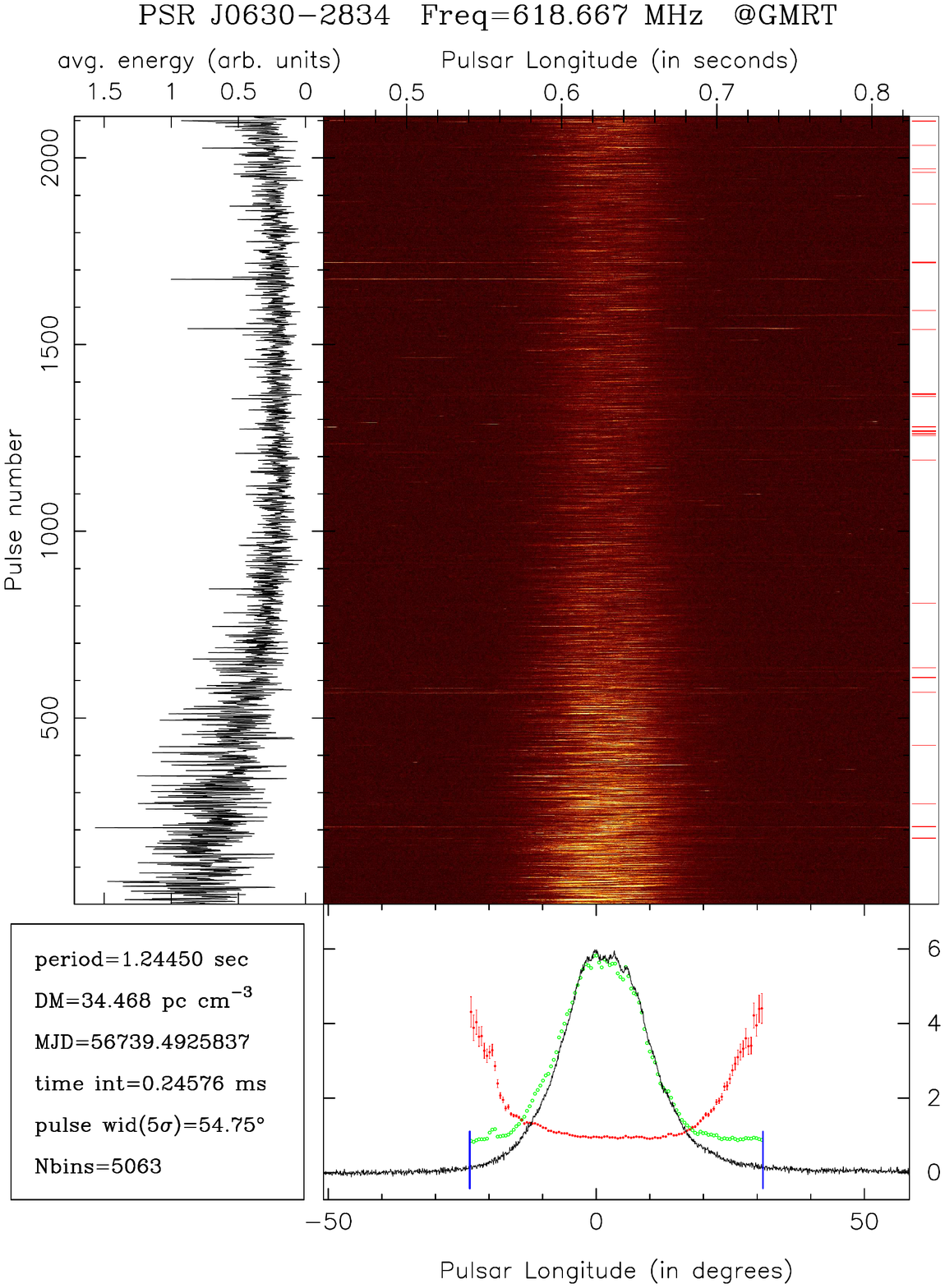}}}\\
\end{tabular}
\caption{The central panel on the left and right plots show single pulse 
stacks of PSR J0630-2834 at 333 and 618 MHz respectively. The red horizontal lines 
on the rightmost strip correspond to pulses affected by RFI. The blue vertical 
bars in the bottom plot correspond to the leading and trailing edge of the pulse 
estimated at 5$\sigma$ level, the black curve corresponds to the average total intensity, 
the green points in the bottom panel are the rms of the intensity fluctuation along each longitude 
and the red points show the modulation index. The black curve in the left panel shows the variation 
of the average single pulse intensity as a function of pulse number. See section~\ref{sec4_2_1} for detail. The figures for all pulsars are available for download (see appendix).} 
\label{figspuls}
\end{center}
\end{figure*}

\begin{figure*}
\begin{center}
\begin{tabular}{cc}
{\mbox{\includegraphics[scale=0.29,angle=0.]{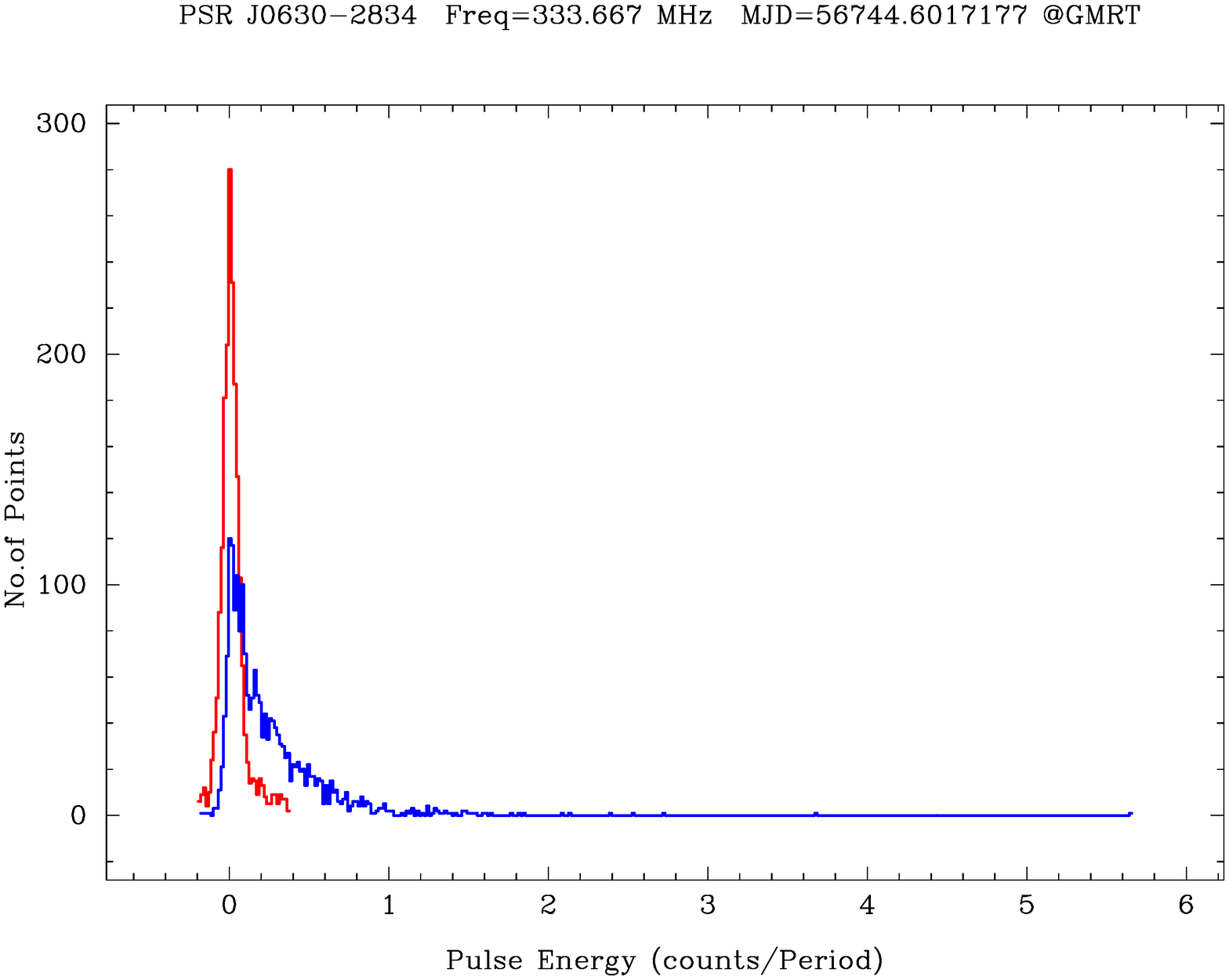}}}& 
{\mbox{\includegraphics[scale=0.29,angle=0.]{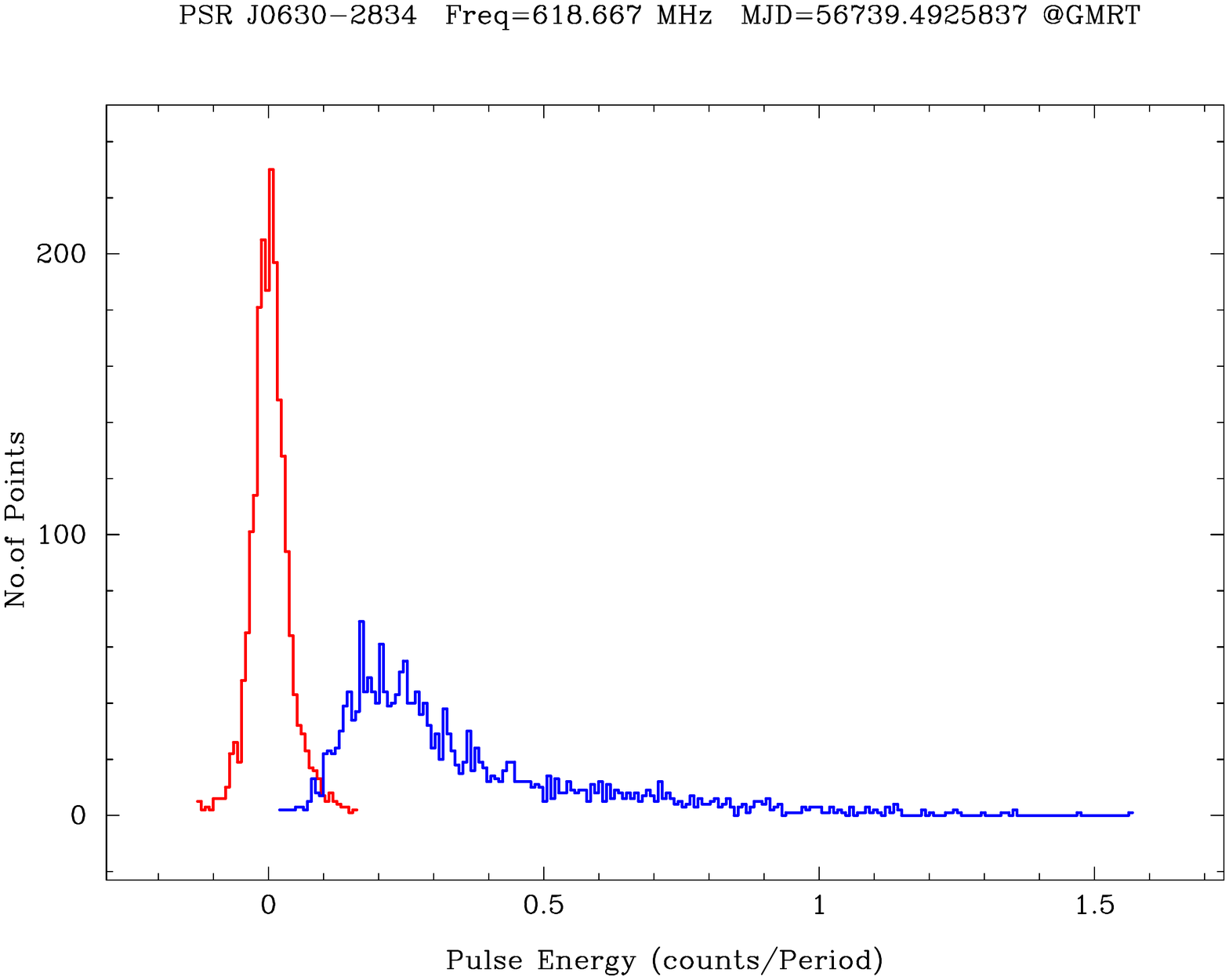}}}\\
\end{tabular}
\caption{The left and right plots show the off-pulse (red curves) and on-pulse 
(blue curves) energy histograms for PSR J0630-2834 
based on the data in figure~\ref{figspuls} for 333 and 618 MHz respectively. See section~\ref{sec4_2_1} for detail. The figures for all pulsars are available for download (see appendix).} 
\label{figspuls1}
\end{center}
\end{figure*}


\begin{figure*}
\begin{center}
\includegraphics[scale=0.70,angle=0.]{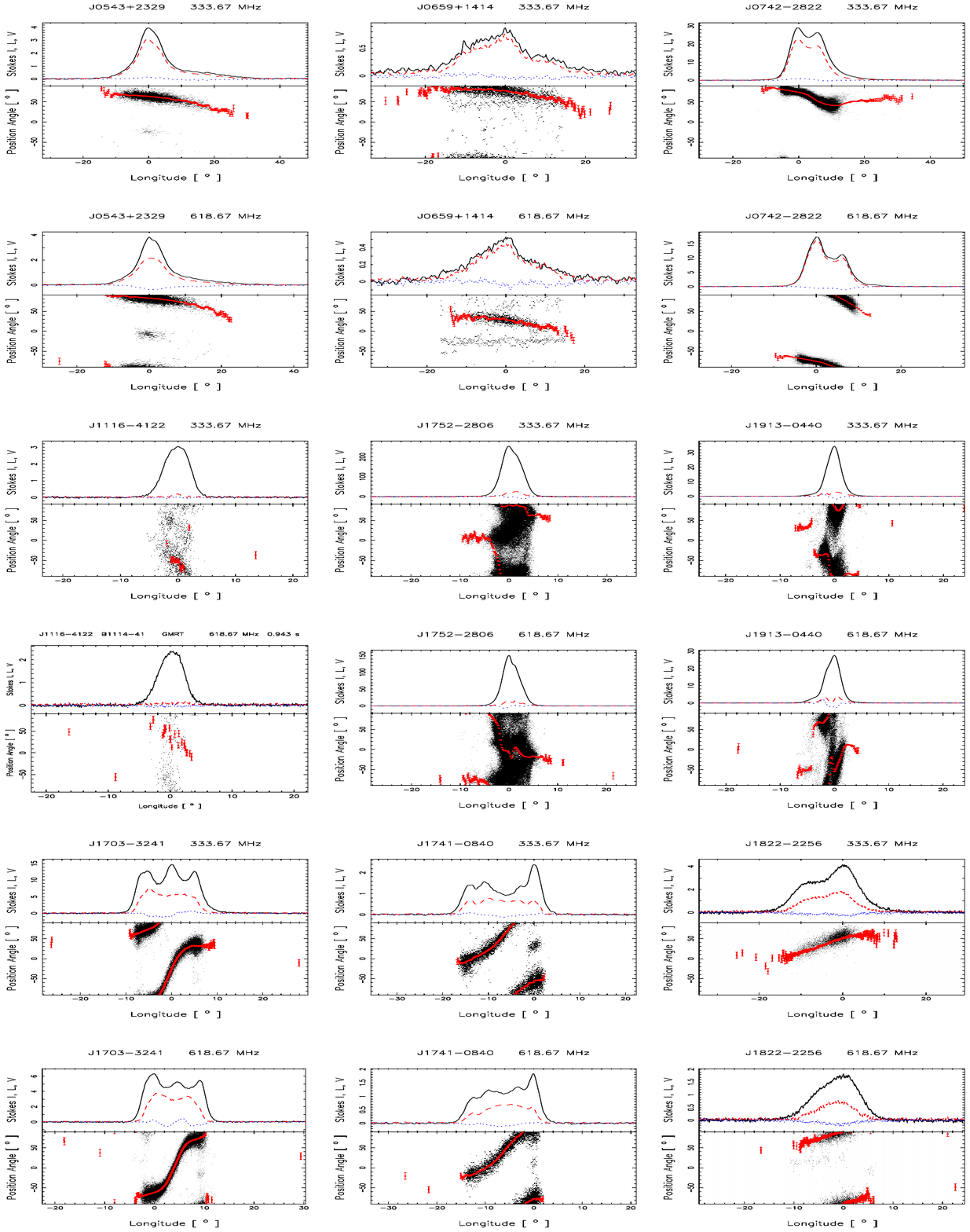}
\caption{Example of single pulse polarization position angle (PPA) histograms for three different ranges of $\dot{E}$, with three pulsars shown for each range at both 333 and 618 MHz. 
The top two panels correspond to $\dot{E} > 10^{34}$~erg~s$^{-1}$, middle two panels correspond to 
$\dot{E} \sim 10^{32}-10^{34}$~erg~s$^{-1}$ and bottom two panels corresponds to $\dot{E} < 10^{31}$~erg~s$^{-1}$. A running mean of 3 bins has
been applied to the data before plotting to increase the S/N of the PPA points. The figures for all pulsars are available for download (see appendix).}
\label{figspols}
\end{center}
\end{figure*}

\subsubsection{\bf Average polarization properties} 
\label{sec4_1_2}
 In table.~\ref{tab3} we list the average degree of linear $\%L = \sum_i L(\phi_i)/I(\phi_i)$,
 circular $\%V = \sum_i V(\phi_i)/I(\phi_i)$ and absolute circular $\%\mid V \mid = \sum_i \mid V(\phi_i)
 \mid/I(\phi_i)$ polarization where the summation is across pulse 
 phase $\phi_i$ and is performed for only statistically significant
 values, i.e. with S/N $>$ 3, for the respective quantities,
 and the noise were estimated from the off-pulse region.  It is to be
 noted that while Stokes $I,Q,U$ and $V$ have errors with gaussian
 distribution, the quantities $\% L$, $\% V$ and $\%\mid V \mid$ have
 non-gaussian error distributions and hence error estimates for these
 quantities cannot be computed using standard error propagation
 \citep{2015ApJ...806..236M}. Instead we took the following approach
 to estimate the errors.  Using the noise rms obtained from the off
 pulse region, we constructed a large number of profiles by varying
 each of the four stoke parameters randomly within the rms value.  For
 each of these profiles we estimated the average $\%L, \%V$ and $\%
 \mid V \mid$ as described above.  The median and the rms values of
 this distribution were used as estimates and errors respectively.
 Figure~\ref{fig1} shows the details of these measurements for one
 pulsar.  In a number of cases, PSR J1648--3256, J1848--1414,
 J1849--0636 and J1921+1948 at 333 MHz and PSR J1739--2903 (main
 pulse), J1757--2421, J1835--1020, J1848--1414 and J1921+1948 at 618
 MHz, there were insufficient polarization measurements above the rms
 cutoff to yield any average polarization values despite the total
 intensity profile having sufficient signal to noise. These are
 examples of extreme depolarization in the pulsar population.

The degree of polarization has been shown to be correlated with pulsar
period $P$, period derivative $\dot{P}$ and their derived parameters
particularly the spindown energy loss $\dot{E} = 4\pi^2 I
\dot{P}P^{-3}$ erg s$^{-1}$ where $I$ is the moment of inertia having
a typical value of 10$^{45}$ g cm$^2$, and characteristic age $\tau =
0.5P \dot{P}^{-1}$ yr (\citealt{1998MmSAI..69.1055V,1998A&A...334..571V}, GL98,
\citealt{2001AJ....122.2001C, 2009ApJS..181..557H,2008MNRAS.391.1210W}, hereafter WJ08).  
The strongest correlation is observed between $\%L$ and $\dot{E}$, with high $\dot{E}$ 
pulsars showing higher degree of linear polarization.  The correlation was a
highlight of the study conducted by WJ08 at 1.4 GHz using 350 pulsars
in the $\dot{E}$ range $\sim 10^{30}-10^{37.5}$ erg s$^{-1}$, where a
distinct but gradual transition is seen around $\dot{E} \sim
10^{34}-10^{35}$ erg s$^{-1}$, above which the degree of polarization
is very high with a mean value of 60\% and below this level it
decreases to 20\%.  A similar correlation in the meterwavelengths has
also been reported by GL98.  We have observed similar correlations in
our data where the dependence of $\%L$ with $P$, $\dot{P}$, $\dot{E}$
and $\tau$ is shown in figure~\ref{fig3}.  It is to be noted that
there exists a spread in the measured values of the degree of
polarization along all $\dot{E}$, however the mean values obtained by
dividing the data along multiple bins in $\dot{E}$ reveal clear trends
in agreement with WJ08.  The minimum of the mean linear polarization
is about 20\% at $\dot{E} \sim 10^{32.3}$ erg s$^{-1}$, which rises
marginally to 30\% below this level but increases significantly to
70\% in the higher $\dot{E}$ range.  Another interesting dependence of
$\%L$ on $\tau$ is observed where younger pulsars ($\tau < 10^6$ yr)
show very high polarizations upto 70\% compared to older ones ($\tau
\gg 10^6$ yr) with $\%L \sim 30\%$.  Our data shows no evidence
for any correlation 
for $\%V$ and $\%\mid V \mid$ to any of these pulsar parameters.

\subsection{Single Pulse Properties}
The observed pulsed emission can be analysed in a number of
ways as illustrated in figures~\ref{figspuls} and~\ref{figspuls1}
for the pulsar J0630$-$2834 and figure~\ref{figspols} at the two observing 
frequencies. The data can be used to study pulsar phenomenon involving 
both single pulse total intensity and single pulse polarization. 

\subsubsection{\bf Single pulse total intensity} 
\label{sec4_2_1}
%
A useful way of plotting single pulse data are single pulse 
stacks where the intensity corresponding to consecutive pulsar period 
is plotted on top of each other as shown in figure~\ref{figspuls}.
The main panel in
the plot represents colour coded contour of single pulse stack of total intensity
$I(j,\phi_i)$ , which corresponds to the $j^{\rm{th}}$ pulse along 
the y-axis and the pulse phase $\phi_i$ along the x-axis. The
baseline levels for each of the single pulses were estimated from a
region of minimum rms in the off-pulse window which was subtracted
from each pulse to create the stack.  In addition the pulses affected
by RFI were identified using a statistical approach, determining the
mean and rms in a off-pulse window for each pulse where the pulses
exceeding five times the median noise identified as outliers, and
shown on the rightmost strip of the figure (horizontal red lines).
The outlier pulses were not included in subsequent analysis.  The
on-pulse window bounded by the longitude range $n_1$ and $n_2$ was
determined as the region above 5 times the rms of the off-pulse window
and marked with blue bordering lines.  The average energy
corresponding to each single pulse was determined as
$\sum_{i=n_1}^{n_2} I(j,\phi_i)/N_{{\rm bins}}$ within the
on-pulse window having $N_{{\rm bins}}$ number of bins, and is shown
as the black curve on the left panel.  The average profile after
rejecting the RFI affected pulses is shown as the black curve in the
lowermost panel.  In addition we have also estimated the fluctuation
of the pulse to pulse intensity along every longitude $\phi_i$ by
measuring the rms of $I(j,\phi_i)$ which is shown as the green points in
the lowermost panel.  The longitude resolved modulation index
$m(\phi_i)$ is shown as the red points with errorbars in the lowermost
panel, and is defined as $m(\phi_i) = \left(\sqrt{<I(j,\phi_i)^2> -
  <I(j,\phi_i)>^2}\right)/<I(j,\phi_i)>$ where the angle brackets
indicate mean values, and the errors were estimated
using Monte Carlo simulations (see \citealt{2012MNRAS.424..843W}).  
The MSPES data set showed a rich variety 
in the distribution of on-pulse energy of single pulse intensity 
which are related to phenomenon like pulsar nulling, 
moding (e.g. \citealt{2007MNRAS.377.1383W}) and interstellar 
scintillation (e.g. \citealt{1977ARA&A..15..479R}). 
This variation can be effectively represented by plotting 
on-pulse and off-pulse energy histogram of single pulses 
(see \citealt{1976MNRAS.176..249R}), an example of which is shown for PSR J0630-2834 in 
figure~\ref{figspuls1}. 
The red histogram corresponds to the off-pulse energy 
histogram which was computed by first finding a off-pulse window in the average profile
which corresponds to a minimum rms regions, and then using the same window to find the 
mean energies of the single pulses.
For purely white noise the  off-pulse histogram should have a normal distribution.
However, as seen in figure~\ref{figspuls1} the distribution, particularly at 333 MHz is asymetric, 
and arises due to a  presence of gain variations leading to systematics in the 
baseline level, and low level RFI which could not be detected through our RFI excision algorithm.
The on-pulse histogram is the blue histogram in figure~\ref{figspuls1}, and was computed 
by finding the single pulse energy in the on-pulse window corresponding to 5$\sigma$ pulse width.  
The on-pulse energy distribution has contribution from the low level RFI and baseline variations as well as
from pulsar single pulse phenomenon associated with nulling, moding and scintillation.
A proper investigation of these phenomenon will need to address the issue of mitigating the low level RFI 
and baseline systematics (one method to eliminate baseline systematics have been devised in MSPESII, appendix A). 
Currently more detailed study of these phenomenon are underway and will be reported elsewhere.
%
%

The next important single pulse phenomenon is subpulse drifting.
The pulsed emission is composed of one or more components called
subpulses which in certain pulsars is seen to exhibit periodic
variation.  This phenomenon is known as subpulse drifting and has been
a subject of considerable interest for understanding the radio
emission mechanism.  The largest study of subpulse drifting has been
conducted by \citet{2006A&A...445..243W, 2007A&A...469..607W} where
187 pulsars were studied and drifting features reported in 68 pulsars
with 42 new detections.  A comprehensive study of the phenomenon of
drifting subpulses using the present dataset has been carried out by
\citeauthor{2016APJ.1..1} (2016, hereafter MSPESII) which is being
presented as an accompanying paper. 
The principal outcomes of the
studies are as follows: we detected drifting features in 39 pulsars at
333 MHz, and 44 pulsars at 618 MHz with a total of 57 pulsars showing
some features of drifting.  The drifting phenomenon was detected for
the first time in 22 pulsars which increased the sample of drifting
pulsars by around 20\% and is one of the largest such studies
conducted. In table~\ref{tab2} pulsars showing drifting are indicated
as ``D'' and the new detections are indicated as ``D$^{\star}$''. 
As demonstrated
in MSPESII, the superior quality of single-pulses in the current study
enabled us to estimate the drifting properties with much higher
significance.

\subsubsection{\bf Single Pulse Polarization} 
\label{sec4_2_2}
Several examples of single pulse polarisation are shown in 
figure~\ref{figspols}.
The plot represents the distribution of the
single pulse phased-resolved PPA (grayscale), with only statistically
significant points exceeding three times the off-pulse noise levels
shown in the plot.  The average PPA is overlaid as a red curve.  
We found around 60\% of the
pulsars in our sample to exhibit PPA histograms with some discernible
points within the pulse window (see table~\ref{tab2}).  The circular polarization of the
single pulses were weak for most pulsars barring a few bright cases,
which will be studied in a future paper.

The PPA histograms exhibit a variety of shapes ranging from simple
S-shaped curves (RVM) to extremely complex structures.  In some cases
the two orthogonal polarization tracks are clearly visible.
The PPA values are determined within
(-90,+90) degrees and at any given longitude there are two distinct
distributions.  A tightly bunched distribution confined to around
10-20 degrees, mostly seen in pulsars with S-shaped PPA tracks, and a
wide spread which sometimes cover the entire window.  Despite a large
variation in shape, a pattern connecting the PPA histogram and the
average degree of linear polarization emerges in our sample.  In
figure~\ref{figspols} we show examples of pulsars with three distinct
PPA behaviour at 333 MHz and 610 MHz: (1) the top two panels show
pulsars with high linear polarization, $\%L \sim 60-70\%$, which are
typically associated with $\dot{E} \geq 10^{34}$ erg s$^{-1}$.  The
single pulse polarization in these cases are close to the average
value and the PPA histograms show tight bunching, sometimes with a
hint of weak orthogonal polarization modes.  (2) The middle two panels
are examples of pulsars with extremely low linear polarization, $\%L <
10\%$, associated with $\dot{E}$ between 10$^{32}$ and 10$^{34}$ erg
s$^{-1}$.  The PPA histograms show chaotic shapes with random spread
within the window.  (3) The bottom two panels are examples of pulsars
with intermediate polarizations, $\%L \sim 30\%$ and $\dot{E} <
10^{32}$ erg s$^{-1}$ cases.  These typify PPA with low spread,
resembling the RVM, and exhibit clear orthogonal polarization modes.
A detailed study of single pulse polarization and how it leads to
depolarization in average profiles will be presented elsewhere.

\section{Summary}
\label{sec5}
In this paper we have described the time-averaged and single pulse
emission properties of the pulsars in MSPES conducted using the GMRT at 333 MHz
and 618 MHz.  These observations were aimed at a systematic and
detailed study of the pulsar radio emission properties.  The
calibrated data sets (section~\ref{sec3}) have been used to estimate
the pulse widths and average linear and circular polarizations in the
pulsar sample (table~\ref{tab3} and discussion in section~\ref{sec4}).
The effect of RFM is clearly demonstrated in the pulse widths with an
estimated power law index of $a\sim$-$0.19$ for the evolution of
widths between 333 MHz and 618 MHz.  The pulse width distribution with
period had a lower bound, with the LBL corresponding to 2.7$^{\circ}P^{-0.5}$,
5.7$P^{-0.5}$ and 6.3$P^{-0.5}$ for $W_{50}$, $W_{10}$ and
$W_{5\sigma}$ respectively at 618 MHz.  The LBL for $W_{50}$ at 1.4 GHz was found
to be 2.45$^{\circ}P^{-0.5}$ by \citet{2012MNRAS.424.1762M} and interpreted as
emission from the narrowest angular structure, mainly the core/conal
component, in a pulse profile.  They invoke the partially screened
vacuum gap (PSG) model \citep{2003A&A...407..315G} of the inner
accelerating region which was initially suggested by
\citet{1975ApJ...196...51R}, and argue that the components in a pulse
profile are related to the sparking discharge in the PSG.  They
further demonstrate that the numerical factor $\sim$2.45$^{\circ}$ in the $W_{50}$
widths can be related to the radius of curvature of non-dipolar
magnetic field in the PSG where dense electron-positron plasma is
created due to the sparking process and finally the pulsar radio
emission arises at altitudes of about 50 stellar radii above the
neutron star surface.  Our result for $W_{50}$ LBL support the findings
of \citet{2012MNRAS.424.1762M}, where the slightly higher numerical
factor 2.7$^{\circ}$ at 618 MHz can be attributed to RFM.  The $W_{10}$ and
$W_{5\sigma}$ LBL however is not connected to the component widths,
instead they measure widths for both the components and the separation
of the components and the LBL suggests the non existence of pulsed
radio emission structures below this level in the pulsar population.
The $P^{-0.5}$ dependence of the widths follows from the nature of the
dipolar open magnetic field lines in the radio emission region. The
physical origin of this bound is currently unclear.
  
We found $\%L$ to be correlated with various pulsar parameters
confirming previous studies.  The degree of correlation varies with
various parameters, e.g. it is much stronger with $P$ than $\dot{P}$.
$\%L$ is seen to be as high as 70\% for pulsars rotating faster than
$\sim$300 milliseconds and about 30\% for periods slower than 400
milliseconds.  The correlation of $\%L$ is also present with both
$\dot{E}$ and $\tau$, where $\%L$ is 70\% for $\dot{E} > 10^{35}$ erg
s$^{-1}$ and $\tau < 10^{5.5}$ yr and $\%L$ is 30\% for $\dot{E} <
10^{34}$ erg s$^{-1}$ and $\tau > 10^{6}$ yr.  The physics of
depolarization in pulsars is still poorly understood, which makes
these results, particularly the transitions between high and low
$\%L$, as important inputs into the various models.  One likely source
of depolarization is the presence of orthogonal and non-orthogonal
polarization modes \citep[e.g.][]{2003ApJ...590..411R}.
Our single pulse polarization data showed that
several pulsars with low $\%L$ exhibit a variety of OPM and non-OPM
distributions compared to high $\%L$ pulsars.  A detailed single pulse
study revealing the relation between polarization properties of
individual pulses and average pulses is essential to understand the
depolarization process.  Additionally, the high quality single pulse
data obtained in our survey showed clear presence of nulling
and subpulse drifting. In an accompanying paper, MSPESII, a
detailed study of drifting subpulses revealed that around 45\% of the
pulsars in our sample exhibit drifting features with 22 pulsars 
in which this phenomenon was detected for the first time
as indicated in table~\ref{tab2}. 

Our data is consistent with the observational evidence that the
coherent radio emission in pulsars originating at heights of 50
stellar radii or below 10\% of the light cylinder
\citep{1991ApJ...370..643B, 1993ApJ...405..285R, 1997MNRAS.288..631K}
which suggests the presence of strong magnetic fields ($\sim 10^{8}$
G) in the emission region.  In such strong magnetic fields the radio
emitting plasma is constrained to move only along the field lines with
all transverse motions suspended. In this specialized condition only
the two-stream instability can develop within the plasma and our
current understanding is that the non-linear growth of the two stream
instability can lead to formation of charged relativistic solitons
emitting coherent curvature radiation in plasma
\citep{2000ApJ...544.1081M, 2004ApJ...600..872G, 2009ApJ...696L.141M,
  2014ApJ...794..105M}.  The supply of the radio emitting plasma from
the inner accelerating region is initiated by a sparking process
\citep[e.g][]{1975ApJ...196...51R} and the drifting subpulse
phenomenon is thought to be associated with the $\vec{E} \times
\vec{B}$ drift, where $\vec{E}$ and $\vec{B}$ are the electric and
magnetic fields in the inner accelerating region.  The detailed study
of the drifting subpulse phenomenon of our data in MSPESII confirms
the presence of the inner accelerating region and favours the PSG
model.  However major challenges are faced when the coherent curvature
radiation theory is used to explain the polarization properties.  The
curvature radiation can excite the ordinary ordinary (O-mode) and
extraordinary (X-mode) modes within the plasma
\citep{2000ApJ...544.1081M, 2004ApJ...600..872G, 2009ApJ...696L.141M,
  2014ApJ...794..105M}.  The X-mode, with the waves polarized
perpendicular to the magnetic field planes, can emerge from the plasma
without suffering any propagation effect. This is supported 
by good observational
evidence where the linear polarization vector emerges from the pulsar 
as X-mode \citep{2001ApJ...549.1111L, 2005MNRAS.364.1397J,
  2015ApJ...804..112R}.  The presence of OPMs on the other hand also
suggests the emergence of the O-modes where the polarization is in the
plane of the magnetic field line.  However, theoretically
considerations expect the O-modes to be heavily damped within the
plasma and unable to emerge (\citealt{1986ApJ...302..120A},\citealt{2014ApJ...794..105M}). 
Also there is no theoretical basis to
understand the presence of circular polarization observed in pulsars.
We aim to carry out a systematic study of identifying the OPM in our
large sample to better understand the polarization properties in pulsars.

The data presented in this paper as well as other studies strongly
suggest that the emission properties depend on pulsar parameters,
particularly $\dot{E}$.  It appears that $\dot{E}$ change leads to a
systematic change in the radio emitting plasma, thereby affecting the
pulsar emission properties.  Our future aim is to use these
observational results and perform further detailed analysis of emission
properties to enhance our understanding of the physics of pulsar radio
emission under the framework of the coherent curvature radio emission
model, with our present work being a significant attempt towards these
goals.

\section*{Acknowledgments}
We would like to thank Late Prof. Janusz Gil for his leadership and
inspiration that has motivated us to start the MSPES project. 
We thank the referee for his comments which helped to improve the paper.
We thank Joanna Rankin, W. Lewandowski and J. Kijak for critical comments 
on the manuscript. We would like to thank the staff of GMRT and NCRA for 
providing valuable support in carrying out this project.  
This work was supported by grants DEC-2012/05/B/ST9/03924 and DEC-2013/09/B/ST9/02177 
of the Polish National Science Centre. This work has 
been supported by Polish National Science Centre grant 
DEC-2011/03/D/ST9/00656 (KK). This work was financed by the Netherlands 
Organisation for Scientific Research (NWO) under project "CleanMachine" 
(614.001.301).

\bibliography{mspes}

\appendix
\begin{center}
{\Large \bf Appendix}
\end{center}
\noindent
We have made the plots and data products from our survey freely available to the user.

\noindent 
Several of the data products for each pulsar have been archived in the website:\\
\url{http://mspes.ia.uz.zgora.pl/}

\noindent
The bulk download of files is also available from:\\
\url{ftp://ftpnkn.ncra.tifr.res.in/dmitra/MSPES/}

\input{mspes_table1.txt}
\input{mspes_table2.txt}

\end{document}

%% file: mspes_table1.txt
\clearpage
\begin{deluxetable}{lccccccccccclccccl}
\rotate
\tabletypesize{\scriptsize}
\tablecolumns{18}
\tablewidth{0pc}
\tablecaption{\small The table gives the sample of pulsars observed. 
The first 8 columns gives various pulsar parameter obtained from the ATNF pulsar catalog. 
In the rotation measure (RM) column the quantities with (NA) are pulsars for which RM was not available in the ATNF catalog. 
Columns 9$-$13 gives single pulse properties for 333 MHz data where we give the number of single pulses N$_{\rm p}$ which are not affected due to RFI, the average peak signal to noise ratio S/N$_{\rm{avg}}$, the percentage of pulses f$_{5\sigma}$ for which the peak signal to noise ratio exceeds the 5 times the off-pulse rms and the percentage of single pulses f$_{\rm{pol}}$ for which single pulse polarization is detected. f$_{\rm{pol}}$ is only quoted when the number of pulses exceeds 100. 
Columns 14$-$18 corresponds to 618 MHz data. 
The letters D in column 13 and 18 stands for pulsars showing drifting features, while D$^*$ represents drifting reported for the first time (see MSPESII).}
\tablehead{
\colhead{}&\colhead{}&
\colhead{}&\colhead{}&
\colhead{}&\colhead{}&
\colhead{}&\colhead{}&
\multicolumn{5}{c}{\underline{333 MHz}}&
\multicolumn{5}{c}{\underline{618 MHz}} \\
     & {Jname} & Period & $\dot{P}_{-15}$ & DM                       & RM                          & $\tau_{6}$ & $\dot{E}_{31}$ & N$_{\rm p}$  & S/N$_{\rm{avg}}$ & f$_{5\sigma}$ & f$_{\rm{pol}}$ &   &  N$_{\rm p}$  & S/N$_{\rm{avg}}$ & f$_{5\sigma}$ & f$_{\rm{pol}}$ &  \\
     &         &  (s) & (s~s$^{-1}$) & ($\frac{\rm{pc}}{\rm{cc}}$) & ($\frac{\rm{rad}}{\rm{m}^2}$) & (yr) & (erg~s$^{-1}$) &   &   & (\%) & (\%) & &  &  & (\%) & (\%) &  }
\startdata
1   &  J0034-0721  &  0.942 & 0.4080 &  10.92 &    9.89 & 36.6  & 1.92    & 2042 & ~13.3 &  ~62 & 54   & D & 1124 & ~~5.7 &  ~40 & 19   & D \\
2   &  J0134-2937  &  0.136 & 0.0784 &  21.81 &   13.00 & 27.7  & 120     & 2180 & ~~3.7 & $<$5 & $<$5 &   & 2144 & ~~3.1 & $<$5 & $<$5 &   \\
3   &  J0151-0635  &  1.464 & 0.4436 &  25.66 &    2.00 & 52.4  & 0.556   & ~615 & ~~5.6 &  ~44 & 11   & D & ~519 & ~~5.5 &  ~40 & 25   & D \\
4   &  J0152-1637  &  0.832 & 1.30   &  11.93 &    2.00 & 10.2  & 8.88    & 2054 & ~10.8 &  ~93 & 39   & D & 2079 & ~~8.7 &  ~88 & 24   & D \\
5   &  J0206-4028  &  0.630 & 1.20   &  12.90 &   -4.00 & 8.33  & 18.9    & 1993 & ~~6.3 &  ~71 &  7   &   & 2080 & ~~3.1 & $<$5 & $<$5 &   \\
    &              &        &        &        &         &       &         &      &       &      &      &   &      &       &      &      &   \\
6   &  J0304+1932  &  1.387 & 1.30   &  15.74 &   -8.30 & 17.0  & 1.91    & 2058 & ~13.1 &  ~81 & 68   & D & 2026 & ~~8.5 &  ~72 & 56   & D \\
7   &  J0452-1759  &  0.548 & 5.75   &  39.90 &   13.80 & 1.51  & 137     & 1868 & ~20.9 &  ~99 & 86   &   & 2158 & ~~8.0 &  ~85 & 33   &   \\
\enddata
\label{tab2}
\end{deluxetable}

%% file: mspes_table2.txt
\clearpage
\begin{deluxetable}{ccccccccccccc}
\rotate
\tabletypesize{\tiny}
\tablecolumns{14}
\tablewidth{0pc}
\tablecaption{\small The table gives the average width and polarization properties. Pulsars with superscript $^{\star}$ are highly scattered pulsars; superscript $^{\dagger}$ are pulsars with interpulses where we have only reported the main pulse emission.
}
\tablehead{
\colhead{}&
\multicolumn{6}{c}{\underline{333 MHz}}&
\multicolumn{6}{c}{\underline{618 MHz}} \\
{\bf PSR } & $\%L$ &$\%V$ & $\mid V \mid$ & $W_{5\sigma}$ &  $W_{10}$ &  $W_{50}$ & $\%L$ &$\%V$ & $\mid V \mid$ & $W_{5\sigma}$ &  $W_{10}$ & $W_{50}$ \\
           &     &    &               &  ($^{\circ}$) &($^{\circ}$) &($^{\circ}$) &     &    &               & ($^{\circ}$)  &($^{\circ}$)&($^{\circ}$)}
\startdata
J0034-0721 & 19.7 $\pm$ 0.2 & 15.8  $\pm$ 0.2 & 15.9  $\pm$ 0.2 & 49.6  $\pm$ 0.1 & 41.6 $\pm$ 0.1 & 22.8 $\pm$ 0.1 & 20.4 $\pm$ 0.8 &  3.9  $\pm$ 0.9 &  7.7  $\pm$ 0.8 & 32.9  $\pm$ 0.1 &  --- $\pm$ --- & 23.4 $\pm$ 0.1 \\ 
J0134-2937 & 70.7 $\pm$ 2.4 & -7.0  $\pm$ 1.9 & 11.0  $\pm$ 1.7 & 31.7  $\pm$ 0.9 &  --- $\pm$ --- & 16.2 $\pm$ 0.6 & 69.5 $\pm$ 2.5 & -15.5 $\pm$ 2.2 & 16.6  $\pm$ 2.1 & 23.9  $\pm$ 0.9 &  --- $\pm$ --- & 18.8 $\pm$ 0.9 \\ 
J0151-0635 & 33.9 $\pm$ 0.9 & -7.0  $\pm$ 1.2 & 11.0  $\pm$ 0.8 & 39.7  $\pm$ 0.1 &  --- $\pm$ --- & 36.6 $\pm$ 0.2 & 38.9 $\pm$ 0.6 & -9.9  $\pm$ 0.7 & 11.9  $\pm$ 0.6 & 37.9  $\pm$ 0.1 &  --- $\pm$ --- & 32.0 $\pm$ 0.1 \\ 
J0152-1637 & 15.5 $\pm$ 0.1 & -1.1  $\pm$ 0.2 & 12.7  $\pm$ 0.1 & 15.0  $\pm$ 0.3 & 11.8 $\pm$ 0.3 &  3.9 $\pm$ 0.3 & 14.4 $\pm$ 0.2 & -4.3  $\pm$ 0.4 & 10.5  $\pm$ 0.2 & 14.4  $\pm$ 0.3 & 11.5 $\pm$ 0.3 &  7.5 $\pm$ 0.3 \\ 
J0206-4028 & 19.7 $\pm$ 0.6 & -14.4 $\pm$ 0.7 & 17.3  $\pm$ 0.5 & 13.8  $\pm$ 0.4 &  8.8 $\pm$ 0.4 &  4.5 $\pm$ 0.2 &  --- $\pm$ --- &  --- $\pm$ ---  &   --- $\pm$ --- &  ---  $\pm$ --- &  --- $\pm$ --- &  --- $\pm$ --- \\ 
           &                &                 &                 &                 &                &                &                &                 &                 &                 &                &                \\
J0304+1932 & 39.5 $\pm$ 0.3 & 11.8  $\pm$ 0.2 & 12.0  $\pm$ 0.2 & 21.9  $\pm$ 0.1 & 19.5 $\pm$ 0.1 & 15.9 $\pm$ 0.1 & 37.6 $\pm$ 0.2 & 11.1  $\pm$ 0.2 & 11.2  $\pm$ 0.2 & 19.7  $\pm$ 0.1 & 17.7 $\pm$ 0.1 & 13.6 $\pm$ 0.1 \\ 
J0452-1759 & 24.3 $\pm$ 0.2 &  3.5  $\pm$ 0.1 &  4.5  $\pm$ 0.1 & 35.8  $\pm$ 0.4 & 26.4 $\pm$ 0.4 & 19.8 $\pm$ 0.4 & 15.8 $\pm$ 0.4 & -2.2  $\pm$ 0.3 &  3.4  $\pm$ 0.2 & 30.6  $\pm$ 0.4 & 24.5 $\pm$ 0.4 & 19.0 $\pm$ 0.4 \\
\enddata
\label{tab3}
\end{deluxetable}